\documentclass{evn2004}
\usepackage{txfonts}
\begin{document}
   \title{Discovering the microJy Radio VLBI Sky via \\ ``Full-beam'' 
          Self-calibration} 
   \author{M.A. Garrett\inst{1}, J.M. Wrobel\inst{2} 
          \and
          R. Morganti\inst{3}
          }
        
\institute{Joint Institute for VLBI in Europe, Postbus 2, 7990
          AA, Dwingeloo, The Netherlands (garrett@jive.nl)\and 
National Radio Astronomy Observatory, P.O. Box O,
Socorro, NM 87801. \and
Netherlands Foundation
          for Research in Astronomy, Postbus 2, 7990 AA, Dwingeloo,
          The Netherlands. }

   \abstract{
     
     We demonstrate that at 1.4 GHz the combined response of sources
     detected serendipitously in deep, wide-field VLBI images is
     sufficient to permit self-calibration techniques to be employed.
     This technique of ``full-beam'' VLBI self-calibration permits
     coherent VLBI observations to be successfully conducted in any
     random direction on the sky, thereby enabling very faint radio
     sources to be detected. Via full-beam self-calibration, Global
     VLBI observations can equal and indeed surpass the level of
     sensitivity achieved by connected arrays. This technique will
     enable large-scale VLBI surveys of the faint radio source
     population to be conducted, in addition to targeted observations
     of faint sources of special interest (e.g.  GRBs, SNe, SNR and
     low-luminosity AGN).
}


\titlerunning{Full-beam VLBI Self-calibration} 

   \maketitle
%

\section{Introduction}

Until recently the study of the sub-mJy and microJy radio sources
at milliarcsecond resolution has been limited by the phase
stability of VLBI arrays. While the technique of phase-referencing is
often used to improve the stability and thus coherence time of VLBI
data, the results are often non-optimal, especially in poor observing
conditions or for reference-target separations that are greater than a
few degrees. In this paper, we present a new calibration technique for
VLBI - {\it full-beam} self-calibration.  In this case, the
self-calibration process is driven by the response of {\it multiple
  faint sources}, {\it i.e.} sources that serendipitously lie close to
the target source, within the confines of the FWHM of a typical VLBI
antenna's primary beam. While individually these faint sources may not
be strong enough for self-calibration techniques to be employed, their
{\it summed response} will often be more than sufficient.  The
technique relies on being able to detect all these faint sources
simultaneously, and thus requires the employment of wide-field VLBI imaging
techniques.  In this paper we demonstrate the feasibility of applying
full-beam self-calibration techniques to (wide-field) VLBI data.

\section{Deep, Wide-field VLBA-GBT Imaging of the Bootes Field in NOAO-N} 

We have recently completed a deep 1.4~GHz VLBA-GBT wide-field survey of
a region located within the NOAO-N Bo\"otes field. The observing
programme employed both traditional external and ``in-beam''
phase-reference techniques (see Garrett, Wrobel \& Morganti 2004, for
the gory details).  Applying wide-field VLBI techniques, a total of 61
sources, selected from a Westerbork Synthesis Radio Telescope (WSRT)
image, were surveyed simultaneously with a range of different
sensitivities and resolution. A total of 9 sources were detected over
a field of 1017~arcmin$^2$ $=$ 0.28~deg$^2$. The inner few arcmins of
the field reaches unprecedented VLBI noise levels of $\sim
9\mu$Jy/beam, rising to $\sim 55\mu$Jy/beam at the edge of the field.
The field and the detections obtained from the full 24 hr data set are shown
in Fig.~1. Each of the VLBI detections has a brightness temperature in
excess of $10^5$~K and morphology that strongly suggests that
their radio emission is powered by AGN processes. For a full scientific
discussion of the nature of the sources, see Garrett, Wrobel \&
Morganti (2004).

   \begin{figure*}
     \centering \vspace{400pt} \includegraphics{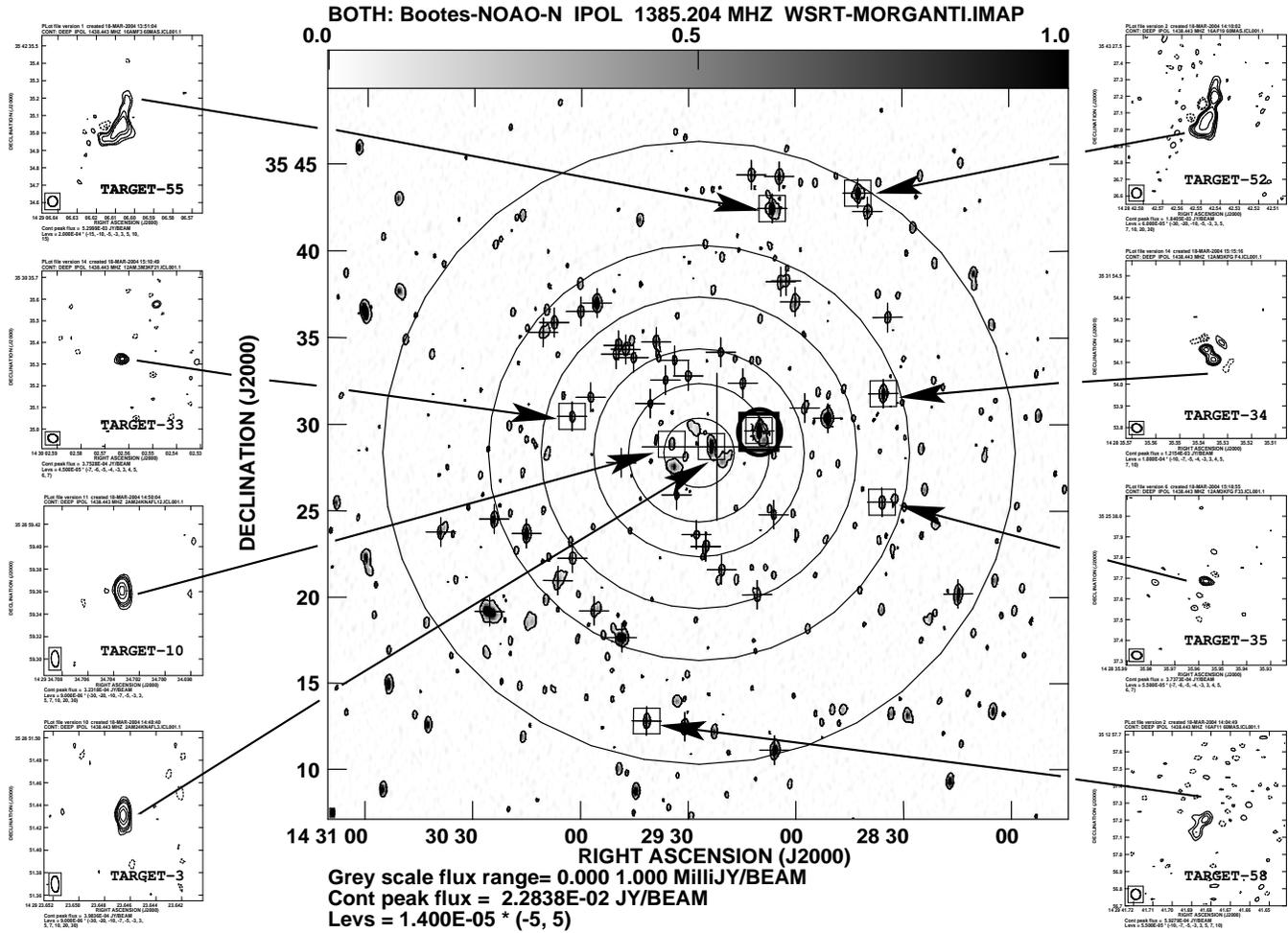}
   \caption{Nine of the compact radio source targets detected simultaneously 
  by VLBI in the Bo\"otes field (contour maps) and boxed in the
  WSRT finding image. The ``in-beam'' calibrator lies $\sim 4$~arcmins
  from the phase centre and is thickly circled and
  boxed.}
   \end{figure*}
   
   The VLBI detection rate for sub-mJy WSRT radio sources in the
   Bo\"otes field is 8$^{+4}_{-5}$\%.  As expected, the VLBI detection
   rate for mJy WSRT sources is higher, 29$^{+11}_{-12}$\%. Our VLBI
   results suggest, that a significant fraction ($\sim 1/3$ ) of the
   sub-mJy AGN radio sources ($> 100$~microJy) are sufficiently compact
   to be detectable with VLBI.


\section{Full-beam Self-calibration} 

The simultaneous detection of several sub-mJy and mJy radio sources in
a single observation, suggests that their combined response may be used
to self-calibrate wide-field VLBI data. We have attempted to
investigate this possibility by self-calibrating a {\it subset} of the
full Bo\"otes VLBI data with all the sources detected in the field,
excluding the bright (20 mJy) ``in-beam'' calibrator (the response of
which was already subtracted from the data as part of the deep field
analysis process). A subset of the data were selected (8 hr of the 24
hr total) with a further constraint that the uv-distance be restricted
to $ < 3$M$\lambda$. This sub-set was chosen in order that the
self-calibration process be tractable on reasonable time-scales,
using a standard (Linux) dual-processor PC.  The uv-data limit was also
required in order to reduce time and bandwidth smearing effects for
sources at the edge of the field. Of the 8 sources detected in the
original analysis, 2 of the fainter sources could not be detected in
the restricted (and thus less sensitive) 3M$\lambda$ data set. However,
the remaining six sources provide a summed (CLEANed) flux density
response of $\sim 20$~mJy across the field.

   \begin{figure*}
     \centering \vspace{210pt} \includegraphics{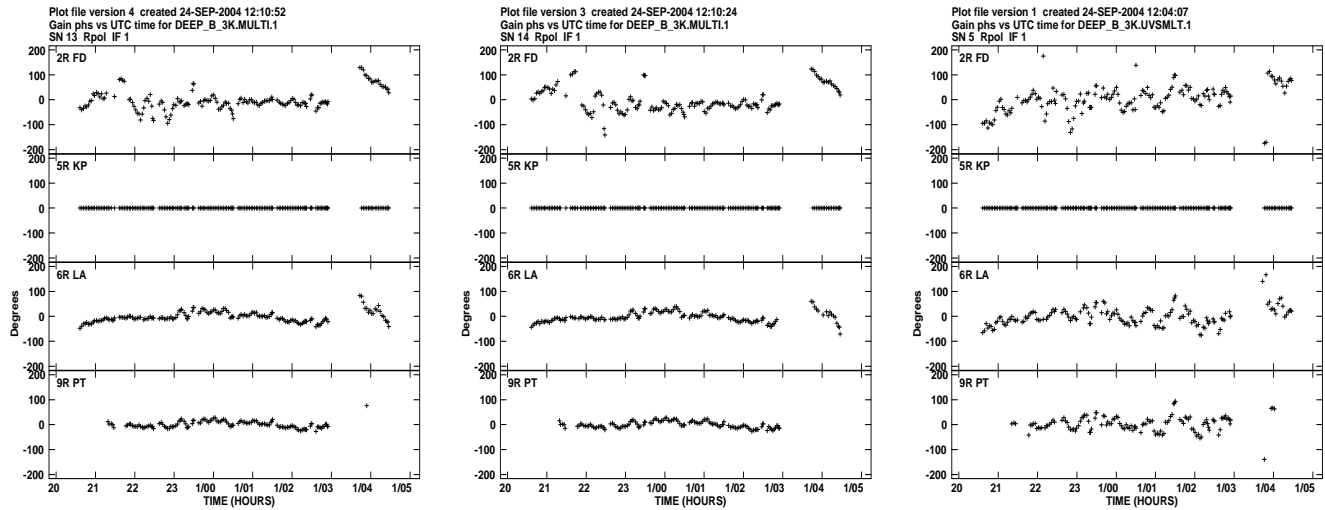}
   \caption{Antenna phase corrections derived from (i) the summed
       ``full-beam'' response
       of the 6 sources in the field - {\it left}, (ii) the response of
       Target 55 alone - {\it middle} \& (iii) the response of 5 sources
       in the field (excluding the brightest, Target 55) - {\it right}.} 
    \end{figure*}
    
    Combined solutions across the full 64~MHz band were obtained with a
    solution interval of $\sim 600$ seconds using the AIPS task, {\sc
      CALIB}. The source model for the self-calibration process,
    included all 6 maps associated with the sources detected in the
    field. Phase solutions were obtained (see Fig.~2, left) with good
    signal-to-noise. The solutions are centred around $0^{\circ}$ phase
    but there are clear deviations from this, reflecting changes in the
    phase corrections derived from the original ``in-beam'' calibrator
    (located close to the centre of the field, see Fig.~1) and those
    derived via the ``full-beam'' technique.
    
    In Figs.~3 we present images of the brightest source (Target 55, $S_{T}
    \sim 13$~mJy) in the field (excluding the $\sim 20$~mJy in-beam
    calibrator) made with the original ``in-beam'' corrections (Fig.~3,
    top left) and ``full-beam'' corrections (also Fig.~3, top right).
    The similarity of the maps suggest that the ``full-beam'' approach
    has worked well.  We also determined phase corrections using the
    response of this bright source (Target 55) alone.  The
    corresponding image is also shown in Fig.~3 (bottom left). Again
    this image is similar to the images obtained via the original
    ``in-beam'' calibrator and ``full beam'' approach.

  \begin{figure}
     \centering \vspace{300pt} \includegraphics{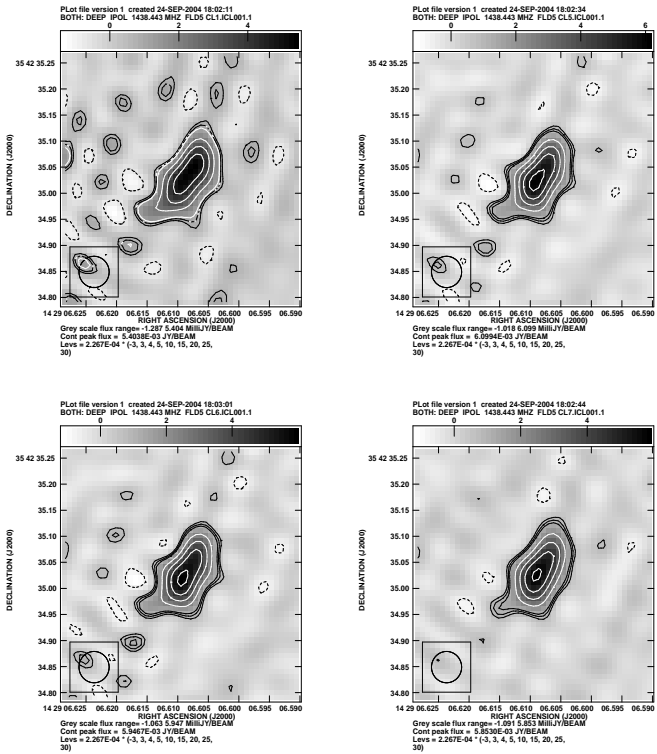}
   \caption{Images of Target 55 with phase corrections derived from: (i)
     the original in-beam calibrator - {\it top left}, (ii) the
     full-beam response of 6 sources in the field - {\it top right},
     (iii) the response of Target 55 alone - {\it bottom left} and (iv)
     the full-beam response of 5 sources in the field, excluding Target
     55 itself - {\it bottom right}. }
    \end{figure}
    
    We also made an additional ``full-beam'' calibrated image, but this
    time the response of the brightest source (Target 55) itself was
    excluded from the self-calibration process. The total summed flux
    density of the remaining 5 sources was $\sim 6.5$~mJy and the
    self-calibration was conducted on a data set for which the response
    of Target 55 had already been subtracted. The phase corrections are
    presented in Fig.~2 (right). These corrections are noisier than the
    previous solutions but are still very similar to those obtained
    from the response of Target 55 only. The similarity is probably due
    to the fact that the phase solutions are weighted towards the
    second brightest source in the field (Target 52) which lies very
    close to Target 55 - at the edge of the field, due north of the
    phase-centre (see Fig.~1).  The solutions were copied back to the
    original data set (that included the response of Target 55) and the
    associated image is presented in Fig.~3 (bottom right). The image
    is similar to the other images presented in Fig.~3, suggesting that
    the full-beam technique has worked well.  

    All images of Target 55 were made with the AIPS task {\sc APCLN}
    using the same CLEAN parameters in each case. Note that corrections
    for ionospheric Faraday rotation and dispersive delay were not made
    to this data set.
    
    We have also made a very preliminary study of the effect of the
    various ``in-beam'' and ``full-beam'' corrections, on the
    detectability of fainter sources in the field, lying far from the
    original ``in-beam'' calibrator and Targets 55 and 52. In
    particular, in Fig.~4 we present images of Target 58 using the
    original ``in-beam'' calibration, and various flavours of
    ``full-beam'' calibration. Target 58 is located far from the field
    centre, almost $0.5^{\circ}$ away from Target 55 and 52, on the
    opposite side of the primary beam (see Fig.~1). The images in
    Fig.~4 show that Target 58 is detected using all flavours of
    full-beam calibration, the maps are all very similar and the 1 mJy
    source is detected with SNR $>$ 13. 

  \begin{figure*}
     \centering \vspace{195pt} \includegraphics{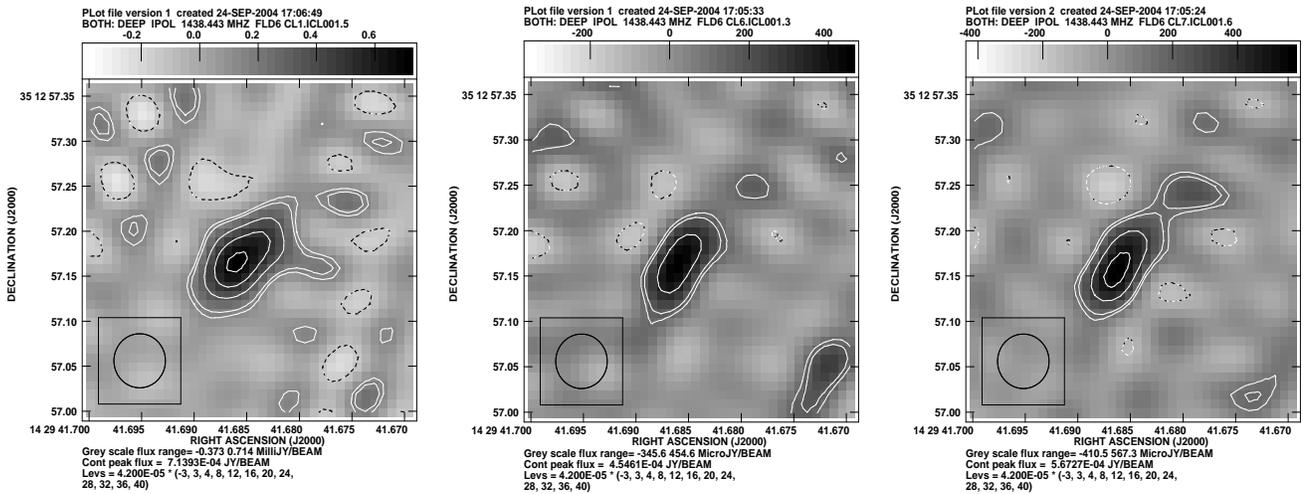}
\caption{Images of the faint radio source, Target 58, 
  with phase corrections derived from: (i) the original in-beam
  calibrator - {\it top left}, (ii) the full-beam response of 6 sources
  in the field - {\it top right}, (iii) the full-beam response of 5
  sources in the field, excluding Target 55 itself - {\it bottom
    left.} }
    \end{figure*}

\section{Conclusions} 

Our study demonstrates that at 1.4 GHz the combined response of sources
detected serendipitously in deep, wide-field VLBI images will often be
sufficient to permit full-beam self-calibration techniques to be
employed. The implication is profound: the application of full-beam
self-calibration permits VLBI observations to be conducted in any
random direction on the sky, thereby enabling large-area, unbiased
surveys of the faint radio source population to be conducted. In
addition, the technique of full-beam calibration can also be used to
improve traditional (nodding) phase reference observations where the
target-calibrator separation is often several degrees and the resulting
images are usually dynamic range limited.  Full-beam VLBI
self-calibration techniques are particularly appropriate for
observations of specific (faint) sources of special interest (e.g.
GRBs, SNe, SNR, low-luminosity AGN {\it etc}).

\begin{figure}
     \centering \vspace{270pt} \includegraphics{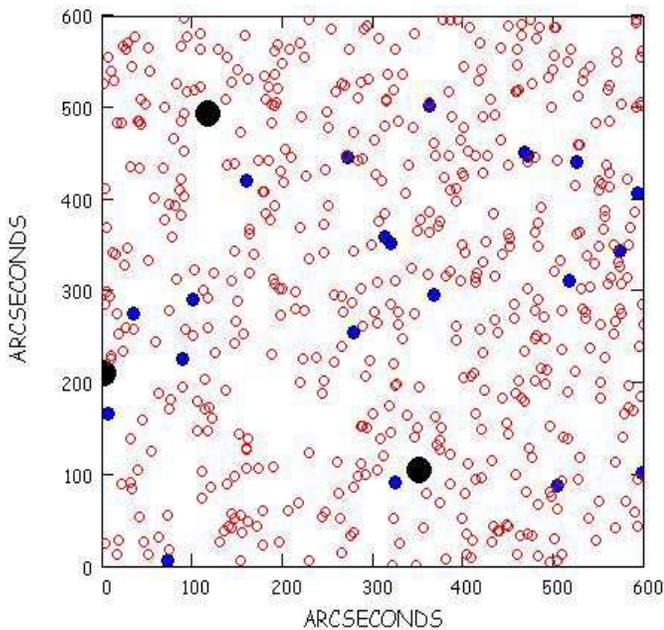}
   \caption{A simulated view of the faint microJy sky as viewed by a
     wide-field VLBI array. There are several hundred potential targets
     in the field of which $\sim 10$\% will be compact enough to be
     detected at milliarcsecond resolution. The summed response of
     these sources will enable full-beam self-calibration to be
     employed in all cases. The radio sources are represented by open
     circles ($S\sim 10-100\mu$Jy), small filled circles ($S\sim 0.1-1$
     mJy) and large filled circles ($S> 1$~mJy). 
}
\end{figure}

As the major VLBI networks upgrade to Mk5 recording systems (permitting
data rates of 1 Gbps), EVN and Global VLBI observations can expect to
reach 1 $\sigma$ rms noise levels of a few microJy/beam. Sources
brighter than $\sim 10\mu$~Jy will be legitimate VLBI targets. Fig.~5
presents a simulated view of the faint microJy sky. Up to 10\% of the
faint sources presented in this figure may be detectable with VLBI and
the summed response of these will enable full-beam calibration to be
employed at frequencies of a few GHz and perhaps as high as 5 GHz for
very deep observations.  For next generation instruments like e-MERLIN,
e-EVN/e-VLBI and the SKA, there will be a wealth of calibration sources
available at almost all frequencies (0.1-25 GHz).

At the levels of sensitivity about to be explored by VLBI, a
comprehensive census of active galaxies associated with sub-mJy radio
sources will be possible, including studies of the optically faint
microJy radio source population.  At microJy noise levels, radio-loud
active galaxies are detectable at the very earliest cosmic epochs, when
the first active galaxies and their energising massive black holes
began to form. In addition, hypernova such as those already detected in
local starburst galaxies (e.g. Arp~220) will be detectable at
cosmological distances, as will GRB after-glows. Full-beam
self-calibration can play an important role in realising these prime
scientific goals.

\begin{acknowledgements}

This research was supported by the European Commission's I3 Programme
``RadioNet", under contract No.\ 505818. NRAO is a facility of the NSF
operated under cooperative agreement by Associated Universities,
Inc. The WSRT is operated by the ASTRON (Netherlands Foundation for
Research in Astronomy) with support from the Netherlands Foundation for
Scientific Research (NWO).

\end{acknowledgements}

\end{document}